\documentclass[twocolumn,aps,superscriptaddress,showpacs]{revtex4}

\usepackage{amssymb}
\usepackage{amsmath}
\usepackage{graphicx}
\usepackage[normalem]{ulem}
\usepackage[dvips]{color}

\begin{document}

\title{Triangular flow in heavy ion collisions
in a multiphase transport model}
\author{Jun Xu}
\affiliation{Cyclotron Institute, Texas A\&M University, College
Station, TX 77843-3366, USA}
\author{Che Ming Ko}
\affiliation{Cyclotron Institute and Department of Physics and
Astronomy, Texas A\&M University, College Station, TX 77843-3366,
USA}

\date{\today}

\begin{abstract}
We have obtained a new set of parameters in a multiphase transport
(AMPT) model that are able to describe both the charged particle
multiplicity density and elliptic flow measured in Au+Au collisions
at center of mass energy $\sqrt{s_{NN}}=200$ GeV at the Relativistic
Heavy Ion Collider (RHIC), although they still give somewhat softer
transverse momentum spectra. We then use the model to predict the
triangular flow due to fluctuations in the initial collision
geometry and study its effect relative to those from other harmonic
components of anisotropic flows on the di-hadron azimuthal
correlations in both central and mid-central collisions.
\end{abstract}


\maketitle

\section{Introduction}
\label{introduction}

Studies of anisotropic azimuthal flows in heavy ion collisions at
RHIC have provided important information on the properties of
produced quark-gluon plasma (QGP)~\cite{Ars05,Adc05,Bac05,Ada05}. In
particular, the large elliptic flow observed in experiments has led
to the conclusion that the produced quark-gluon plasma is strongly
interacting as it can only be explained in the hydrodynamic model
with a very small
viscosity~\cite{teaney01,huovinen02,hirano02,rom07} or in the
transport model with parton scattering cross sections much larger
than those given by the perturbative QCD~\cite{Lin02b,molnar02}.
With the large parton scattering cross section, the transport model
has also been able to describe the hexadecupole flow measured at
RHIC~\cite{Chen04}. More recently, the importance of the triangular
flow, which originates from fluctuations in the initial collision
geometry~\cite{Pau10}, has been pointed out in Ref.~\cite{Alv10a}.
Unlike the elliptic flow, the triangular flow is less sensitive to
the centrality or the impact parameter of the
collision~\cite{Alv10a,Alv10b,Pet10}. Also, a study based on the
viscous hydrodynamics has shown that the viscosity in the
quark-gluon plasma has a larger effect on the triangular flow than
the elliptic flow in relativistic heavy ion collisions~\cite{Sch11}.
Furthermore, it was suggested in Ref.~\cite{Alv10a} and later shown
in Ref.~\cite{Xu11a} in a multiphase transport (AMPT) model, which
includes both initial partonic and final hadronic scatterings, that
the triangular flow may play an important role in the away-side
double-peak structure seen in the di-hadron azimuthal correlations
at RHIC. In the present paper, we extend the study of
Ref.~\cite{Xu11a} to investigate more quantitatively the triangular
flow in heavy ion collisions at RHIC by adjusting the parameters in
the AMPT model, particularly the parton scattering cross section, to
fit more recent experimental data such as the charged particle
multiplicity density and transverse momentum spectra as well as
their elliptic flow. We find that the resulting triangular flow has
smaller values than those shown in Ref.~\cite{Xu11a} but still has
an appreciable effect on the di-hadron azimuthal correlations as
found in Ref.~\cite{Xu11a}.

This paper is organized as follows. In Sec.~\ref{ampt}, we review
the AMPT model and discuss the parameters in the model that are
relevant to the present study. In Sec.~\ref{multiplicity}, we
describe the results on the charged particle multiplicity density
and transverse momentum spectra in heavy ion collisions at RHIC and
their comparisons with experimental data. We then show in
Sec.~\ref{flow} the calculated charged particle elliptic flow in
comparison with the experimental measurements as well as the
predicted triangular flow. In Sec.~\ref{dihadron}, we study the
effect of anisotropic flows, particularly that of triangular flow,
on the di-hadron azimuthal correlations. In Sec.~\ref{eta}, we
discuss the specific viscosity in the initial partonic matter
produced in heavy ion collisions. Finally, a summary is given in
Sec.~\ref{summary}.

\section{The AMPT model}\label{ampt}

The AMPT model is a hybrid model~\cite{Lin05} with the initial
particle distributions generated by the heavy ion jet interaction
generator (HIJING) model~\cite{Xnw91}. In the default version, the
jet quenching in the HIJING model is replaced in the AMPT model by
explicitly taking into account the scattering of mini-jet partons
via the Zhang's parton cascade (ZPC) model~\cite{Zha98}. These
partons are recombined with their parent strings after the
scattering, which are then converted to hadrons using the Lund
string fragmentation model. In the version of string melting, all
hadrons produced from the string fragmentation in the HIJING model
are converted to their valence quarks and antiquarks, whose
evolution in time and space is modeled by the ZPC model. After the
end of their scatterings, quarks and aniquarks are converted to
hadrons via a spatial coalescence model. In both versions of the
AMPT model, the scatterings among hadrons are described by a
relativistic transport (ART) model~\cite{Bal95}.

\begin{table}[h]
\caption{{\protect\small Values of parameters for the Lund string
fragmentation and the parton scattering cross section in previous
studies (A) and in the present work (B).}} \label{tab}
\begin{tabular}{ccccccc}
\hline\hline
 & $a$ & \quad $b$ (GeV$^{-2}$) \quad & $\alpha_s$ & \quad $\mu$ (fm$^{-1}$)  \\
\hline
$$ A~~~  &  2.2 & 0.5 & 0.47 & 1.8 \\
$$ B~~~  &  0.5 & 0.9 & 0.33 & 3.2 \\
\hline\hline
\end{tabular}%
\end{table}

In previous studies, it was found that the multiplicity of charged
particles measured in heavy ion collisions at RHIC could be well
described by the default AMPT model with modified values of $a$ and
$b$~\cite{Lin01} in the Lund string fragmentation function
$f(z)\propto z^{-1} (1-z)^a \exp (-b~m_{\perp}^2/z)$, where $z$ is
the light-cone momentum fraction of the produced hadron of
transverse mass $m_\perp$ with respect to that of the fragmenting
string. To describe the measured elliptic flow required, on the
other hand, the AMPT model with string melting together with a
larger parton scattering cross section $\sigma\approx 9\pi
\alpha_s^2/(2\mu^2)$, where $\alpha_s$ is the QCD coupling constant
and $\mu$ is the screening mass of a gluon in the QGP, than that
given in the perturbative QCD. Values for these parameters are given
in the first row (A) of Table~\ref{tab}. In these
studies~\cite{Chen04,Lin02a}, the elliptic flow was calculated with
respect to the theoretical reaction plane, corresponding to the
second-order event plane $\Psi_2=0$. As a result, this might have
underestimated the elliptic flow that is determined with non-zero
$\Psi_2$ to take into account the fluctuations in the initial
collision geometry as in the experimental analysis. We have recently
shown, however, that both the charged particle multiplicity density
and elliptic flow measured in heavy ion collisions at the Large
Hadron Collider (LHC) can be described by the AMPT model with string
melting by using the default $a$ and $b$ parameters in the HIJING
model together with a smaller QCD coupling constant and a larger
screening mass as given in the second row (B) in
Table~\ref{tab}~\cite{Xu11b}. We note that the parton scattering
cross sections for parameter sets A and B are about 10 mb and 1.5
mb, respectively. Although the latter has a smaller value, it has a
more isotropic angular distribution.

\section{charged particle multiplicity density and transverse momentum spectra at RHIC}
\label{multiplicity}

\begin{figure}[h]
\centerline{\includegraphics[scale=0.8]{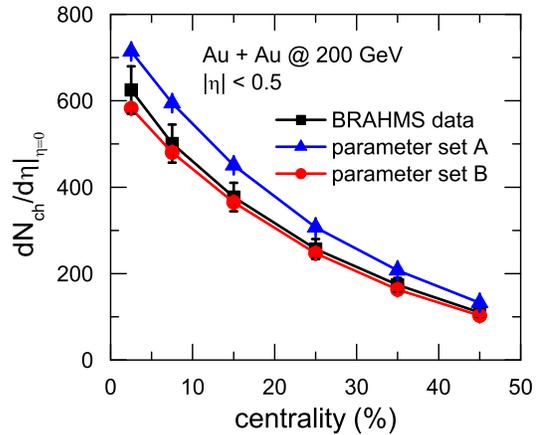}} \caption{(Color
online) Centrality dependence of the multiplicity density of
mid-pseudorapidity charged particles produced in Au+Au collisions at
$\sqrt{s_{NN}}=200$ GeV from the AMPT model with string melting
using parameter sets A (filled triangles) and B (filled circles).
The BRAHMS data shown by filled squares are from Ref.~\cite{Bea02}.}
\label{mul}
\end{figure}

We first show in Fig.~\ref{mul} the centrality dependence of the
multiplicity density of mid-pseudorapidity charged particles in
Au+Au collisions at $\sqrt{s_{NN}}=200$ GeV obtained from the AMPT
model with string melting for the two parameter sets in
Table~\ref{tab} and compare them with the experimental data from the
BRAHMS Collaboration~\cite{Bea02}. For the relation between
the centrality $c$ and the impact parameter b, we use the empirical
formula $c=\pi\text{b}^2/\sigma_{\rm in}$~\cite{Bro02} with the
nucleus-nucleus total inelastic cross section $\sigma_{\rm in}
\approx 705$ fm$^2$ calculated from the
Glauber model.  It is seen that the multiplicity
densities from the parameter set A shown by filled triangles are
larger than the BRAHMS data shown by filled
squares at all centrality bins, while those from the parameter set
B, which are given by filled circles, are consistent with the BRAHMS
data.

\begin{figure}[h]
\centerline{\includegraphics[scale=0.8]{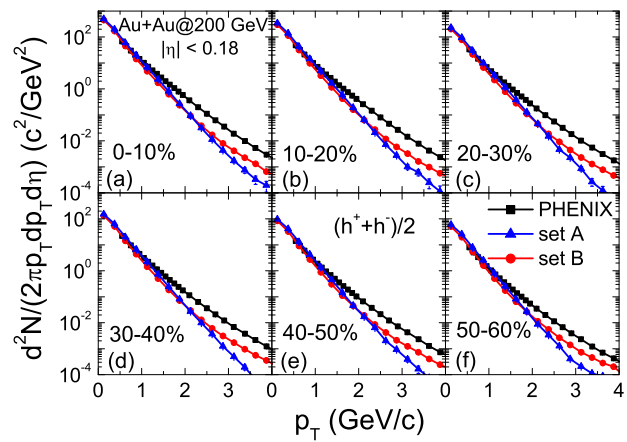}} \caption{(Color
online) Transverse momentum spectra of mid-pseudorapidity
($|\eta|<0.18$) charged particles by using parameter sets A (filled
triangles) and B (filled circles) in Au+Au collisions at
$\sqrt{s_{NN}}=200$ GeV. The PHENIX data shown by filled
squares are taken from Ref.~\cite{Adl04}.} \label{ptdis}
\end{figure}

Figure~\ref{ptdis} shows the transverse momentum ($p_T$) spectra of
mid-pseudorapidity charged particles in different centrality bins
($0-60\%$) from parameter sets A (filled triangles) and B (filled
circles). It is seen that both parameter sets describe reasonably
the experimental data from the PHENIX
Collaboration~\cite{Adl04} at low $p_T$. At high
$p_T$, the parameter set B gives a larger yield than the parameter
set A as a result of smaller energy loss of high-$p_T$ particles
when the particle density is lower and the parton scattering cross
section is smaller as for the parameter set B. As in the previous
study~\cite{Lin05}, the $p_T$ spectra from both parameter sets are
softer than the experimental data, and this is due to the small
current quark masses used in the AMPT model so that partons are less
affected by the radial flow effect.

\section{Charged particle anisotropic flows at RHIC}\label{flow}

\begin{figure}[h]
\centerline{\includegraphics[scale=0.8]{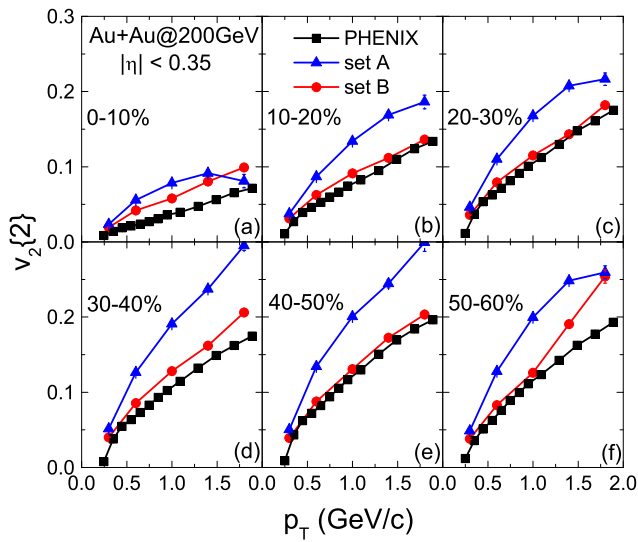}} \caption{(Color
online) Transverse momentum dependence of the elliptic flow of
mid-pseudorapidity ($|\eta|<0.35$) charged particles by using
parameter sets A (filled triangles) and B (filled circles) in Au+Au
collisions at $\sqrt{s_{NN}}=200$ GeV from the two-particle
cumulant method. The PHENIX data shown by filled squares are taken
from Ref.~\cite{Afa09}.} \label{v2pt}
\end{figure}

In Fig.~\ref{v2pt}, we compare the transverse momentum dependence of
the elliptic flow of mid-pseudorapidity charged particles in
different centrality bins ($0-60\%$) of same collisions from the two
parameter sets with the experimental data from the PHENIX
Collaboration~\cite{Afa09}. In both theoretical calculations and
experimental analyses, the elliptic flow is
determined using the two-particle cumulant method~\cite{Wan91,Bor01}
\begin{equation}
v_2\{2\} = \sqrt{\langle \cos(2\Delta\phi) \rangle },
\end{equation}
where $\Delta\phi$ is the azimuthal angular difference between
particle pairs within the same event and $\langle\cdots\rangle$
denotes average over all possible pairs. As the nonflow effect is
overestimated at higher $p_T$ in the AMPT model as shown in
Ref.~\cite{Xu11b}, we only compare the anisotropic flows at lower
$p_T$ where the nonflow effect is small. It is seen that for all
centralities, the elliptic flow is larger for the parameter set A
than for the parameter set B. This originates from two effects.
First, the larger string tension in parameter set A leads to a
larger number of initial particles, which results in a larger
pressure and thus a larger anisotropic flow. Second and more
importantly, the larger parton scattering cross section in parameter
set A converts more efficiently the initial spatial anisotropy to
the final momentum anisotropy. As for the charged particle
multiplicity density and transverse momentum spectra, the parameter
set B gives a much better description of the measured elliptic flow
than the parameter set A.

\begin{figure}[h]
\centerline{\includegraphics[scale=0.8]{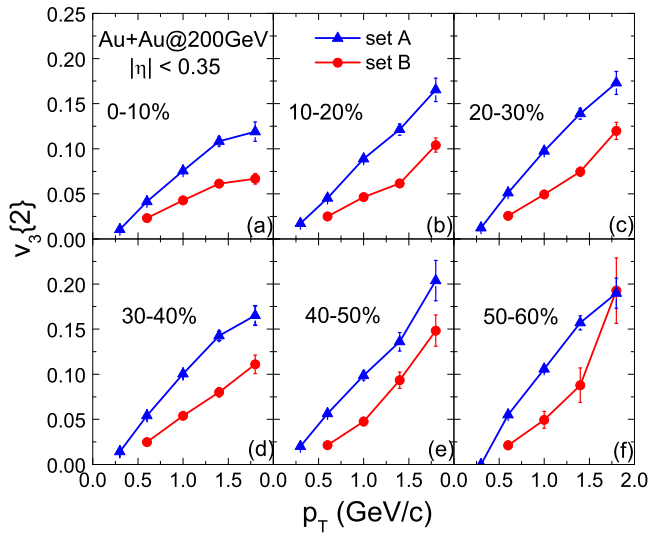}} \caption{(Color
online) Transverse momentum dependence of the triangular flow of
mid-pseudorapidity ($|\eta|<0.35$) charged particles by using
parameter sets A (filled triangles) and B (filled circles) in Au+Au
collisions at $\sqrt{s_{NN}}=200$ GeV from the two-particle cumulant
method.} \label{v3pt}
\end{figure}

We have also studied the transverse momentum dependence of the
triangular flow of mid-pseudorapidity charged particles in different
centrality bins ($0-60\%$) in Au+Au collisions at
$\sqrt{s_{NN}}=200$ GeV. Results obtained from the two-particle
cumulant method, i.e., $v_3\{2\} = \sqrt{\langle \cos(3\Delta\phi)
\rangle}$, are shown in Fig.~\ref{v3pt} using both parameter sets A
(filled triangles) and B (filled circles). Compared with the
elliptic flow shown in Fig.~\ref{v2pt}, the triangular flow is
smaller and less dependent on the centrality. Similar to the case of
elliptic flow, the triangular flow is larger for the parameter set A
than for the parameter set B. Since the parameter set B has been
shown to give a better description of the charged particle
multiplicity density, transverse momentum spectra, and elliptic
flow, we believe that it would also give a more reliable prediction
for the triangular flow. We note that the magnitude of the
triangular flow from the parameter set B is similar to that from the
$(3 + 1)d$ viscous hydrodynamic model~\cite{Sch11} with the specific
viscosity of the value $0.08$ and the
transport+hydrodynamics hybrid approach~\cite{Pet10}
but a little larger than that in Ref.~\cite{Alv10a} based on the
AMPT model.

\begin{figure}[h]
\centerline{\includegraphics[scale=0.8]{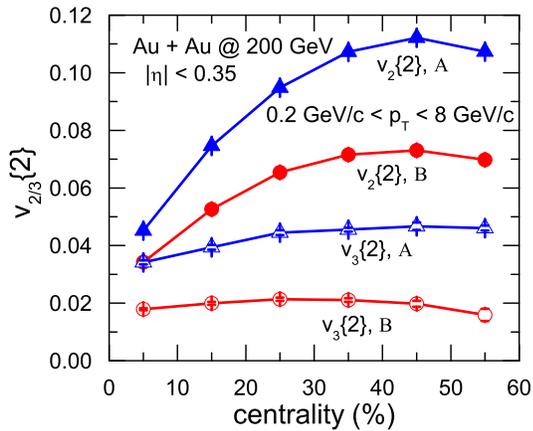}} \caption{(Color
online) Centrality dependence of the elliptic flow (filled symbols)
and the triangular flow (open symbols) from parameter sets A
(triangles) and B (circles) in Au+Au collisions at
$\sqrt{s_{NN}}=200$ GeV for mid-pseudorapidity ($|\eta|<0.35$)
charged particles in the transverse momentum window $0.2$ GeV/c
$<p_T<8$ GeV/c from the two-particle cumulant method.}
\label{v23cen}
\end{figure}

Figure~\ref{v23cen} displays the centrality dependence of the
elliptic flow (filled symbols) and the triangular flow (open
symbols) of mid-pseudorapidity charged particles in the transverse
momentum window $0.2$ GeV/c $<p_T<8.0$ GeV/c from the two-particle
cumulant method. Compared to the elliptic flow, the triangular flow
shows a much weaker centrality dependence. As for the
$p_T$-dependent differential flows, the momentum-integrated flows
are larger for the parameter set A than for the parameter set B.
Both elliptic and triangular flows increase with increasing
centrality at small centralities but saturate or decrease at large
centralities. Again, results from the parameter set B are similar to
those from the transport+hydrodynamics hybrid
approach~\cite{Pet10}.

\section{Di-hadron azimuthal correlations}
\label{dihadron}

\begin{figure}[h]
\centerline{\includegraphics[scale=0.8]{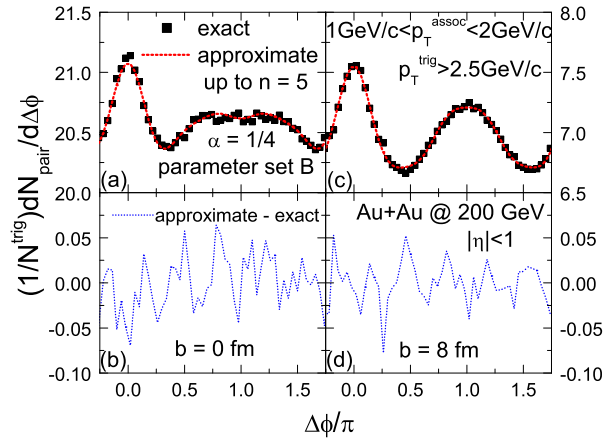}} \caption{(Color
online) Comparison of the exact azimuthal correlations per trigger
particle with the approximate ones defined in Eq.~(\ref{raw}) up to
order $n=5$ in Au+Au collisions at $\sqrt{s_{NN}}=200$ GeV for
impact parameters $\text{b}=0$ fm ((a) and (b)) and $\text{b}=8$ fm
((c) and (d)) from the parameter set B.} \label{comB}
\end{figure}

The effect of anisotropic flows on the di-hadron azimuthal
correlations is a topic of great current
interest~\cite{Alv10a,Mgl11,Xu11a,Aga10,Luz11} and can be studied by
considering following approximate azimuthal correlations:
\begin{eqnarray}\label{raw}
\left\langle\frac{dN_{\rm pair}}{d\Delta\phi}\right\rangle_\text{e} &=&
\frac{1}{2\pi}[\langle N^{\rm trig}N^{\rm assoc}\rangle_\text{e}\notag\\
&+& 2 \sum_{n=1}^{+\infty} \langle N^{\rm trig}N^{\rm assoc}
v_n^{\rm trig} v_n^{\rm assoc}
\rangle_\text{e} \cos(n\Delta\phi)],\notag\\
\end{eqnarray}
where $\langle\cdots\rangle_\text{e}$ denotes average over all
events, $N^{\rm trig}$ and $N^{\rm assoc}$ are numbers of trigger
and associated particles, and $v_n^{\rm trig}$ and $v_n^{\rm assoc}$
are, respectively, their $n$th-order anisotropic flows. The
$n$th-order anisotropic flow is calculated with respect to the
corresponding event plane defined by
\begin{equation}\label{ep}
\psi_n = \frac{1}{n} \arctan \frac{\langle
p_T^\alpha\sin(n\phi)\rangle}{\langle p_T^\alpha\cos(n\phi)\rangle},
\end{equation}
where $\phi$ is the azimuthal angle and $p_T^\alpha$ is the weight
factor. As in Ref.~\cite{Mgl11}, we take the transverse momenta of
trigger particles to be $p_T^{\rm trig}>2.5~{\rm GeV}/c$,
corresponding mostly from the fragmentation of energetic jets
produced in initial hard scatterings that have not interacted much
with the produced medium. For the associated particles, their
transverse momenta are taken in the window $1~{\rm GeV}/c <p_T^{\rm
assoc}<2~{\rm GeV}/c$ as in Ref.~\cite{Mgl11}, and they are medium
particles after the passage of initially produced back-to-back jet
pairs. At mid-pseudorapidity ($|\eta|<1$), the numbers of trigger
particles and associated particles from the AMPT model are,
respectively, $1.336$ and $128.6$ for $\text{b}=0$ fm, and $0.565$
and $42.3$ for $\text{b}=8$ fm. We first show in the upper panels of
Fig.~\ref{comB} by filled squares the exact di-hadron azimuthal
correlations per trigger particle calculated from all possible pairs
of trigger particles and associated particles from the AMPT model
for the two impact parameters $\text{b}=0$ fm and $\text{b}=8$ fm,
corresponding to central and mid-central collisions, respectively.
It is seen that for both impact parameters there is a peak around
$\Delta\phi=0$ at the near side of trigger particles, while at their
away side, i.e., around $\Delta\phi=\pi$, there is a broad structure
for $\text{b}=0$ fm but a pronounced peak for $\text{b}=8$ fm. As
shown by dashed lines, the approximate di-hadron azimuthal
correlations calculated with Eq.~(\ref{raw}) using $\alpha=1/4$ in
determining the event plane (Eq.~(\ref{ep})) reproduce very well the
exact ones, and their difference, shown in the lower panels of
Fig.~\ref{comB}, is smaller than the residual correlations after
subtracting the contributions from anisotropic flows as discussed
below.

\begin{figure}[h]
\centerline{\includegraphics[scale=0.8]{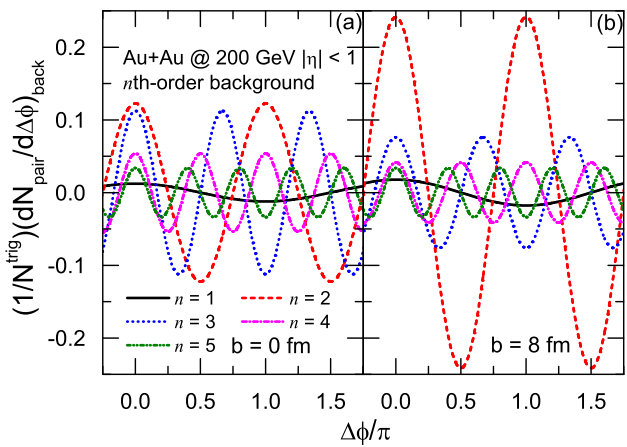}} \caption{(Color
online) Di-hadron azimuthal correlations per trigger particle from
anisotropic flows up to order $n=5$ in Au+Au collisions at
$\sqrt{s_{NN}}=200$ GeV for impact parameters $\text{b}=0$ fm (left
panel) and $\text{b}=8$ fm (right panel) from the parameter set B.}
\label{background}
\end{figure}

As in Ref.~\cite{Xu11a}, we evaluate the contributions to di-hadron
azimuthal correlations from anisotropic flows of various orders by
replacing the event average of products with the product of event
averages in the approximate correlations given in
Eq.~(\ref{raw}), i.e.,
\begin{eqnarray}
&&\left(\frac{dN_{\rm pair}}{d\Delta\phi}\right)_\text{back} =
\frac{1}{2\pi}[\langle N^{\rm trig}\rangle_e \langle N^{\rm assoc}\rangle_\text{e}\notag\\
&&\quad+ 2 \sum_{n=1}^{+\infty} \langle N^{\rm trig} v_n^{\rm trig}
\rangle_\text{e} \langle N^{\rm assoc} v_n^{\rm assoc}
\rangle_\text{e} \cos(n\Delta\phi)], \label{bg}
\end{eqnarray}
where the first term is a constant independent of the azimuthal
angle $\phi$, and the other terms are the contributions from
anisotropic flows. In Fig.~\ref{background}, we show the di-hadron
azimuthal correlations per trigger particle from anisotropic flows
up to order $n=5$. It is seen that the elliptic flow, which has a
peak at $\Delta\phi=\pi$, has the largest contribution for
$\text{b}=8$ fm, while for $\text{b}=0$ fm the contribution from the
triangular flow, which has peaks at $\Delta\phi=2\pi/3$ and $4\pi/3$
in the di-hadron azimuthal correlations, is equally important.
Furthermore, the triangular flow is seen to give a larger
contribution to the di-hadron azimuthal correlations per trigger
particle for $\text{b}=0$ fm than for $\text{b}=8$ fm as a result of
the larger number of associated particles, although it has a larger
value for $\text{b}=8$ fm than for $\text{b}=0$ fm as shown in
Sec.~\ref{flow}. As to the contributions from higher-order flows
$v_4$ and $v_5$, they are, on the other hand, relatively small for
both $\text{b}=0$ fm and $\text{b}=8$ fm.

\begin{figure}[h]
\centerline{\includegraphics[scale=0.7]{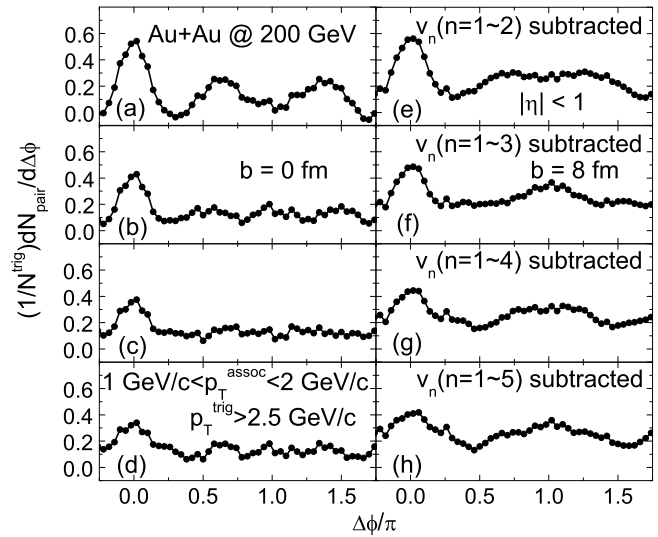}} \caption{(Color
online) Di-hadron correlations per trigger particle after
subtracting background correlations up to different orders in Au+Au
collisions at $\sqrt{s_{NN}}=200$ GeV for
$\text{b}=0$ fm ((a), (b), (c) and (d)) and $\text{b}=8$ fm ((e),
(f), (g) and (h)) from the parameter set B.} \label{dihadronB}
\end{figure}

The residual di-hadron correlations per trigger particle after
subtracting both the constant term in Eq.~(\ref{bg}) and the
contributions due to anisotropic flows (Fig.~\ref{background}) from
the exact azimuthal correlations (upper panels in Fig.~\ref{comB})
are shown in Fig.~\ref{dihadronB}. It is seen from
Figs.~\ref{dihadronB}(a) and (e) that the away-side double-peak
structure is much stronger for $\text{b}=0$ fm than for $\text{b}=8$
fm after subtracting the contribution from the direct flow as well
as the large contribution from the elliptic flow. Subtracting also
the contribution from the triangular flow as shown in
Figs.~\ref{dihadronB}(b) and (f) changes the away-side double peaks
in the di-hadron azimuthal correlations to essentially flat
correlations for $\text{b}=0$ fm but to a single peak for
$\text{b}=8$ fm, while the near-side peak is reduced in both cases.
As expected from the small contributions due to higher-order flows
shown in Fig.~\ref{background}, further subtraction of the
contributions from higher-order anisotropic flows does not change
much the shape of the di-hadron correlations as shown in
Figs.~\ref{dihadronB}(c) and (g) as well as in
Figs.~\ref{dihadronB}(d) and (h). The residual di-hadron azimuthal
correlations per trigger particle after the subtraction of the
contributions from anisotropic flows are then the correlations
induced by initially produced back-to-back jet pairs. Our results
indicate that the away-side jets essentially disappear in the
produced medium in collisions at $\text{b}=0$ fm but are still
visible in collisions at $\text{b}=8$ fm, consistent with the
observed larger jet quenching in central than in mid-central
collisions. The away-side double-peak structure in
Figs.~\ref{dihadronB}(a) and (e) after subtracting the contribution
from the elliptic flow is thus from the combined effects of the
away-side jets and the higher-order anisotropic flows, particularly
the triangular flow. The weaker away-side double-peak structure for
$\text{b}=8$ fm in Fig.~\ref{dihadronB} (e) than for $\text{b}=0$ fm
in Fig.~\ref{dihadronB} (a) then reflects the effect due to the
remnant of away-side jets and the relatively smaller contribution
per trigger particle from the triangular flow shown in
Fig.~\ref{background}.

It is interesting to compare present results based on the parameter
set B with those in Refs.~\cite{Xu11a} and \cite{Mgl11} based on the
parameter set A for $\text{b}=8$ fm and $\text{b}=0$ fm,
respectively. For the case of $\text{b}=8$ fm, the away-side
double-peak structure is weaker for the parameter set B than for the
parameter set A, and the reasons for this are similar to those that
cause its weakening in comparison to the case of $\text{b}=0$ fm
based on the same parameter set. For the case of $\text{b}=0$ fm,
the conclusion that the away-side double-peak structure in the
results from the parameter set B is mainly due to the triangular
flow is different from that in Ref.~\cite{Mgl11} based on the
parameter set A, which seems to indicate that other effects such as
jet deflections or Mach cone shock waves are also relevant.

\section{Specific viscosity}\label{eta}

\begin{figure}[h]
\centerline{\includegraphics[scale=0.8]{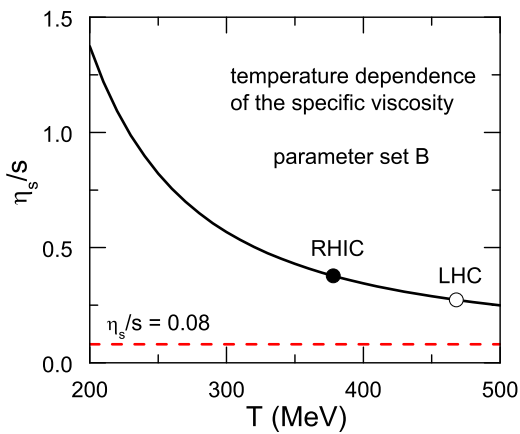}} \caption{(Color
online) Temperature dependence of the specific viscosity in the
partonic matter from the parameter set B.} \label{etas}
\end{figure}

With the success of the parameter set B in describing measured
elliptic flow in heavy ion collisions at both RHIC and LHC, it is of
interest to estimate the specific viscosity in the partonic matter
formed in these collisions. In the kinetic theory, the shear
viscosity is given by $\eta_s=4\langle p\rangle/(15\sigma_{\rm tr})$
in terms of the parton mean momentum $\langle p\rangle$ and the
parton transport or viscosity cross section $\sigma_{\rm tr}=\int
dtd\sigma/dt(1-\cos^2\theta)$, where $t$ is the standard Mandelstam
variable for four-momentum transfer and $d\sigma/dt\approx
9\pi\alpha_s^2/[2(t-\mu^2)^2]$ is the differential cross section
used in the AMPT model. By assuming that the partonic matter only
consists of non-interacting massless up and down quarks as in the
AMPT model, we have $\langle p\rangle=3T$ and the entropy density
$s=(\epsilon+P)/T=4\epsilon/(3T)=96T^3/\pi^2$ with $T$ being the
temperature of the partonic matter. The specific viscosity, i.e.,
the ratio between the shear viscosity and the entropy density, is
then~\cite{Xu11b}
\begin{eqnarray}\label{vis}
\eta_s/s\approx
\frac{3\pi}{40\alpha_s^2}\frac{1}{\left(9+\frac{\mu^2}{T^2}\right)\ln\left(\frac{18+\mu^2/T^2}{\mu^2/T^2}\right)-18}.
\end{eqnarray}
The temperature dependence of the specific viscosity obtained from
the parameter set B is shown in Fig.~\ref{etas}, and it shows that
the specific viscosity decreases with increasing temperature. With
the energy density of the baryon-free quark and antiquark matter
given by $\epsilon=72T^4/\pi^2$, the initial temperature $T$ is
found to be about $378$ MeV in Au+Au collisions at
$\sqrt{s_{NN}}=200$ GeV and $468$ MeV in Pb+Pb collisions at
$\sqrt{s_{NN}}=2.76$ TeV from the average energy density of
mid-rapidity partons at their average formation time~\cite{Xu11b}.
The specific viscosity is then about $0.377$ at RHIC energy and
$0.273$ at LHC energy as indicated, respectively, by the solid and
open circles in Fig.~\ref{etas}.  The values are thus similar at
RHIC and LHC, although both are much larger than the lower bound
$\eta_s/s\approx 0.08$ predicted by the AdS/CFT
correspondence~\cite{Kov05}.

\section{Summary}
\label{summary}

Using the default values for the parameters in the Lund string
fragmentation function and a smaller but more isotropic parton
scattering cross section than previously used in the AMPT model for
heavy ion collisions at RHIC, we have obtained a good description of
both the charged particle multiplicity density and the elliptic flow
measured in Au+Au collisions at $\sqrt{s_{NN}}=200$ GeV, although
the transverse momentum spectra are still softer than the
experimental results. With these constrained parameters, the
magnitude of the triangular flow in these collisions has been
predicted. We have also studied the di-hadron azimuthal correlations
triggered by energetic hadrons at both impact parameters of
$\text{b}=0$ and $8$ fm and found that
the double-peak structure at the away side of triggered particles,
which is seen after subtracting the background contributions due to
the elliptic flow, is largely due to the triangular flow. However,
the residual correlations shown in our study after the subtraction
of the flow contribution might still contain the contribution
from flow fluctuations besides the nonflow contribution that we are
interested in~\cite{Yi11}. It will be of great interest to find a
method that can disentangle the nonflow contribution from that due
to flow fluctuations.

We have also estimated the specific viscosity in the initial
partonic matter and found that it is much larger than the lower
bound predicted by the AdS/CFT correspondence and is thus different
from the values extracted with the viscous hydrodynamic model. The
different conclusions from the hydrodynamic model and the AMPT model
might come from the fact that a constant specific viscosity is used
in the former model while a constant total cross section is used in
the latter one. Including the temperature dependence of the local
screening mass in the evaluation of the parton scattering cross
section in the AMPT model as in Ref.~\cite{Zha10} may help to better
understand the different results from the transport model and the
hydrodynamic model.

\begin{acknowledgments}
We thank Fuqiang Wang and You Zhou for helpful comments. This work
was supported in part by the U.S. National Science Foundation under
Grant No. PHY-0758115, the US Department of Energy under Contract
No. DE-FG02-10ER41682, and the Welch Foundation under Grant No.
A-1358.
\end{acknowledgments}

\end{document}